\documentclass[sigconf,nonacm]{acmart}

\AtBeginDocument{%
  \providecommand\BibTeX{{%
    \normalfont B\kern-0.5em{\scshape i\kern-0.25em b}\kern-0.8em\TeX}}}

\setcopyright{none}

\PassOptionsToPackage{table,xcdraw}{xcolor}
\usepackage{colortbl}
\usepackage[utf8]{inputenc}
\usepackage{wrapfig}
\usepackage{graphicx}
\usepackage[toc,page]{appendix}
\usepackage[acronym]{glossaries}
\newacronym{llm}{LLM}{large language model}
\graphicspath{{figures/}}

\usepackage{stfloats}
\usepackage{acmart-taps}

\copyrightyear{2024}
\acmYear{2024}
\setcopyright{rightsretained}
\acmConference[IUI '24]{29th International Conference on Intelligent User Interfaces}{March 18--21, 2024}{Greenville, SC, USA}
\acmBooktitle{29th International Conference on Intelligent User Interfaces (IUI '24), March 18--21, 2024, Greenville, SC, USA}
\acmDOI{10.1145/3640543.3645198}
\acmISBN{979-8-4007-0508-3/24/03}

\begin{document}

\title{Take It, Leave It, or Fix It: Measuring Productivity and Trust in Human-AI Collaboration}
\author{Crystal Qian}
\email{cjqian@google.com}
\orcid{0001-7716-7245}
\affiliation{%
  \institution{Google Research}
  \city{Cambridge}
  \state{MA}
  \country{USA}
 }
\affiliation{%
  \institution{Massachusetts Institute of Technology}
  \city{Cambridge}
  \state{MA}
  \country{USA}}
  
\author{James Wexler}
\email{jwexler@google.com}
\orcid{0009-0006-8105-6998}
\affiliation{%
  \institution{Google Research}
  \city{Cambridge}
  \state{MA}
  \country{USA}}

\begin{abstract}Although recent developments in generative AI have greatly enhanced the capabilities of conversational agents such as Google's Gemini or OpenAI's ChatGPT, it's unclear whether the usage of these agents aids users across various contexts. To better understand how access to conversational AI affects productivity and trust, we conducted a mixed-methods, task-based user study, observing 76 software engineers (N=76) as they completed a programming exam with and without access to Google's Bard.\footnote{At the time of this study in July 2023, Google's generative artificial intelligence chatbot was named \textit{Bard}. This has since been renamed to \textit{Gemini} \cite{Hsiao2024}.} Effects on performance, efficiency, satisfaction, and trust vary depending on user expertise, question type (open-ended "solve" vs. definitive "search" questions), and measurement type (demonstrated vs. self-reported). Our findings include evidence of automation complacency, increased reliance on the AI over the course of the task, and increased performance for novices on “solve”-type questions when using the AI. We discuss common behaviors, design recommendations, and impact considerations to improve collaborations with conversational AI.
\end{abstract}




\maketitle
\begin{sloppypar}
\section{Introduction}\label{intro}

Recent advancements in generative artificial intelligence (AI) have the potential to enhance productivity across domains such as medicine \cite{Li2024}, research \cite{Salvagno2023, Lund2023}, and technology \cite{Rice2024}. In the context of software development, conversational AI such as Google’s Gemini and Open AI’s ChatGPT can generate code, and Github’s Copilot can autocomplete code. However, it’s unclear whether these systems strictly improve productivity. State-of-the-art agents suffer from imperfect accuracy and biases \cite{Oviedo2023,Wach2023}, and humans have demonstrated cognitive biases such as automation bias and effort substitution when using LLM-based systems such as ChatGPT and Copilot \cite{Madi2023, Barke2023, Vai2022, Noy2023}. Furthermore, behaviors and outcomes may  depend on the expertise or literacy of the user; novices have been found to be less discerning of automation errors \cite{Oviedo2023}.

The opportunity remains that many developer tasks can benefit from an open-ended conversational AI interface, such as brainstorming ideas, answering questions \cite{Lund2023}, debugging errors, and addressing subjective topics such as translation \cite{Weisz2022} or coding conventions. To better understand developer interactions with conversational AI, we conducted a user study with 76 software engineers (N=76) at a Google as they completed an occupation-specific programming language exam with and without assistance from a conversational AI agent, Google's Bard. 


We pose the following research questions in this setting:

\begin{itemize}
    \item \textbf{RQ1, Effects on productivity}: How does usage of conversational AI affect productivity?
    \item \textbf{RQ2, Behaviors of trust}: How do users demonstrate trust in conversational AI?
\end{itemize}

To evaluate the value-add of conversational AI on productivity in \textbf{RQ1}, our experimental design randomizes ordering of access to conversational AI within-participant, comparing productivity when adding access to Bard to productivity when adding access to traditional resources. To evaluate \textbf{RQ2}, we construct an action space of trusting and distrusting behaviors. Across both productivity and trust constructs, we consider behavioral and self-reported measures. We also explore how effects may vary across levels of user expertise, constructing expertise rankings using a rich database of company-internal engineering statistics and self-reported survey responses. 

We find that participants of all expertise levels increasingly depend on conversational AI over the course of the exam, despite mixed results on measured and perceived productivity. Novices are particularly influenced by these systems, which could imply opportunities for skill equalization. However, this can be a cause for concern, as generative models may be more likely to propagate misleading information compared to traditional forms of decision support tools with fixed outputs \cite{Wach2023}. We also find evidence of incongruity between users' anticipated and demonstrated behaviors, suggesting that users are not fully cognizant of their interactions with these systems.

This paper contributes the following empirical evidence:
\begin{itemize}
    \item \textbf{Access to AI can affect productivity and perceived productivity in different directions:} Users  perceive \nobreak increased productivity and efficiency when using the AI, despite spending more time on the task.
    \item \textbf{Expertise and context matter:} AI usage can improve \nobreak performance, particularly for novices on open-ended tasks.
    \item \textbf{Users increasingly rely on AI over the course of the task:}  Participants of all expertise levels increasingly depend on the AI, despite reporting less trust in the AI.
    \item \textbf{Experts distrust, and distrust imperfectly:} Relative to novices, experts are more likely to reject the AI. This can punish performance, as experts  are less likely to use the AI to correct mistakes.
    \item \textbf{Usage of AI reduces cognitive load:} Participants \nobreak substitute effort to the AI and report reduced cognitive load.
\end{itemize}

We also discuss common behaviors, design implications, and ethical considerations for more beneficial intelligent systems. This work aims to advance the understanding of productivity and trust formation in using conversational AI. 

\section{Related Work}\label{relatedwork}

\subsection{Usage of conversational agents} Since the release of ChatGPT in 2022, there has been increased interest in evaluating the performance of conversational agents. ChatGPT and GPT-4 can perform sufficiently well on a diverse range of analytical NLP tasks (e.g. sentiment analysis, question answering), but accuracy decreases as the difficulty of the task increases \cite{Kocon2023}. Despite this, these agents are increasingly adopted in professional \cite{Noy2023} and academic \cite{Lund2023, Salvagno2023, Susnjak2022, Rice2024} settings, and are used to inform high-impact decision making around topics such as safety \cite{Oviedo2023} and healthcare \cite{Li2024, Xie2019}. 

\subsection{Variation in interactions by user ability} Not all users interact equally with these systems; populations with lower literacy and education have been found to have higher risk of consuming unreliable content by conversational agents \cite{Oviedo2023}. There are arguments as to why both experts and novices may be more or less amenable to automated assistance. Experts may reject systems due to a rational allocation strategy, deciding not to delegate tasks to an external system if the expert’s internal trustworthiness is high \cite{Lewandowsky2000, Mozannar2023}. On the other hand, novices may exhibit an overestimated belief in their ability (Dunning-Krueger effect \cite{Kruger1999}) and also reject automated assistance \cite{Schaffer2019}. Perceived expertise may be more \nobreak relevant than expertise in trust formation; self-confidence can causally relate to automation usage \cite{Lee1994, Cai2022, Hemmer2023}. Users of \nobreak automated systems across all ability levels have exhibited self-reported, \nobreak implicit, and explicit propensity to trust automation without prior interactions or adequate evidence about its \nobreak capabilities \cite{Merritt2019}, and exhibit higher tolerance for AI misfires \cite{Kapania2022}.

\subsection{Variation in interactions by context} 
On a writing task, ChatGPT usage increased performance for low performers and velocity for high performers \cite{Noy2023}. However, on a programming task, Copilot did not increase the success rate of solving programming tasks or reduce task completion time, as \nobreak developers spent more time validating generated outputs \cite{Vai2022}. \nobreak Writers substituted effort by directly copy-pasting outputs from ChatGPT without verification \cite{Noy2023}, and developers directed less visual attention to Copilot-generated code, despite its quality being comparable to human-written code \cite{Madi2023}.  Behaviors of \nobreak automation bias and effort substitution have also been found in automation environments such as manufacturing and aviation technology \cite{Lewandowsky2000, Wickens2015, Para2010, Zhang2023}. Effort substitution may be the optimal strategy if automated systems can perform perfectly. However, LLM-based systems can regurgitate or hallucinate potentially misleading information \cite{Azamfirei2023, Salvagno2023, Weisz2021}, and there is no one-size-fits-all solution on how to present information optimally in conversational systems to mitigate misinterpretation \cite{Yeh2022, Sun2022, Cau2023}. Despite that user trust is affected by the accuracy of ML systems \cite{Yin2019, Kahr2023}, users are willing to accept help from imperfect assistants \cite{Weisz2021, Prabhudesai2023}, and imperfect assistance can nonetheless improve performance \cite{Weisz2022}.

\section{Experiment Design}\label{design}

\subsection{Procedure}\label{procedure}
We invited a random sample of 1,400 US-based, full-time software engineers at Google to participate in our study. Of the 220 who responded, 96 were accepted given our screening requirement that they had written and submitted code within the last 6 months. 79 respondents completed informed consent \nobreak paperwork and scheduled study sessions, and 76 respondents completed the study (2 dropouts and 1 timeout, a 4$\%$ attrition rate).

Following a pre-task survey, each participant joined a 1-on-1, virtual, hour-long study with a moderator; they completed a 10-question multiple-choice online exam while sharing their screen and thinking aloud. Finally, they completed a post-task assessment. 

\subsection{Task design}\label{task}

\textbf{Exam:} We chose a modified exam format for this task \cite{Susnjak2022, Gilson2023} and calibrated the number of questions across 8 pilot sessions. The ten multiple-choice, single-answer questions on the exam come from a company-internal \nobreak “readability” exam on the Java programming language, which is one component of the process undergone by software engineers to obtain a Java readability certification.\footnote{This certification allows software engineers to submit code without requiring additional Java-specific review. There are also exams for other languages. We chose Java because it is a popular language \cite{Weisz2022} used commonly across our company, which yielded a broader distribution of expertise given our random sample of participants.} The exam tests understanding of the company's coding standards, which is documented in the company's internally- and externally- published Java Style Guide. Coding conventions can be subjective; the style guide is written to ensure consistency across the company. Often, multiple answers in the exam can compile and are technically valid, but one of the answers is more correct than others according to the company’s Style Guide. This is stated in the introduction to the exam: \textit{“There may be multiple correct answers.. Please choose the best one in line with the [Company] Java Readability Style Guide.”
}
We verified that Bard could complete this task independently with reasonable performance; when directly given the exam questions and corresponding multiple-choice options as inputs, Bard could answer 7.5 out of 10 questions correctly on average.\footnote{We repeated trial runs and varied levels of priming and prompt engineering (e.g. providing context such as "\textit{According to the [Company] Style Guide \ldots}").}\textsuperscript{,}\footnote{Note that the task was built in the Qualtrics survey platform, which disabled copy-pasting of multiple-choice answers by default. This inadvertently required participants to exert more effort if they wanted to apply this direct copy-paste strategy.}

\textbf{Format:} We divided the ten exam questions into two sections of five questions each: a "Bard-First" and a "Bard-Last" section. Section order and question order were randomized within-subject. Participants had two passes at each question. During the first pass per question in the “Bard-First” section, participants had access to Bard only. After selecting an answer, they had the option of modifying their answer with access to any non-LLM-based resources of their choice (e.g. documentation, search engines), loosely categorized as "Book" resources. During the first pass in the "Bard-Last" section, participants had access to Book resources. In the second pass, participants could \textit{add Bard}, meaning they could modify their answer using Bard. Our intervention was \textit{access} to these resources; participants did not have to use them. By allowing participants two passes at each question, we isolated the effect of adding access to Bard or Book in the second pass. To help users calibrate trust in the AI, we displayed a feedback screen after the second pass showing the correct answer and explanation \cite{Brunsen2021}.

\begin{figure}[ht]
        \centering
        \includegraphics[width=.45\textwidth]{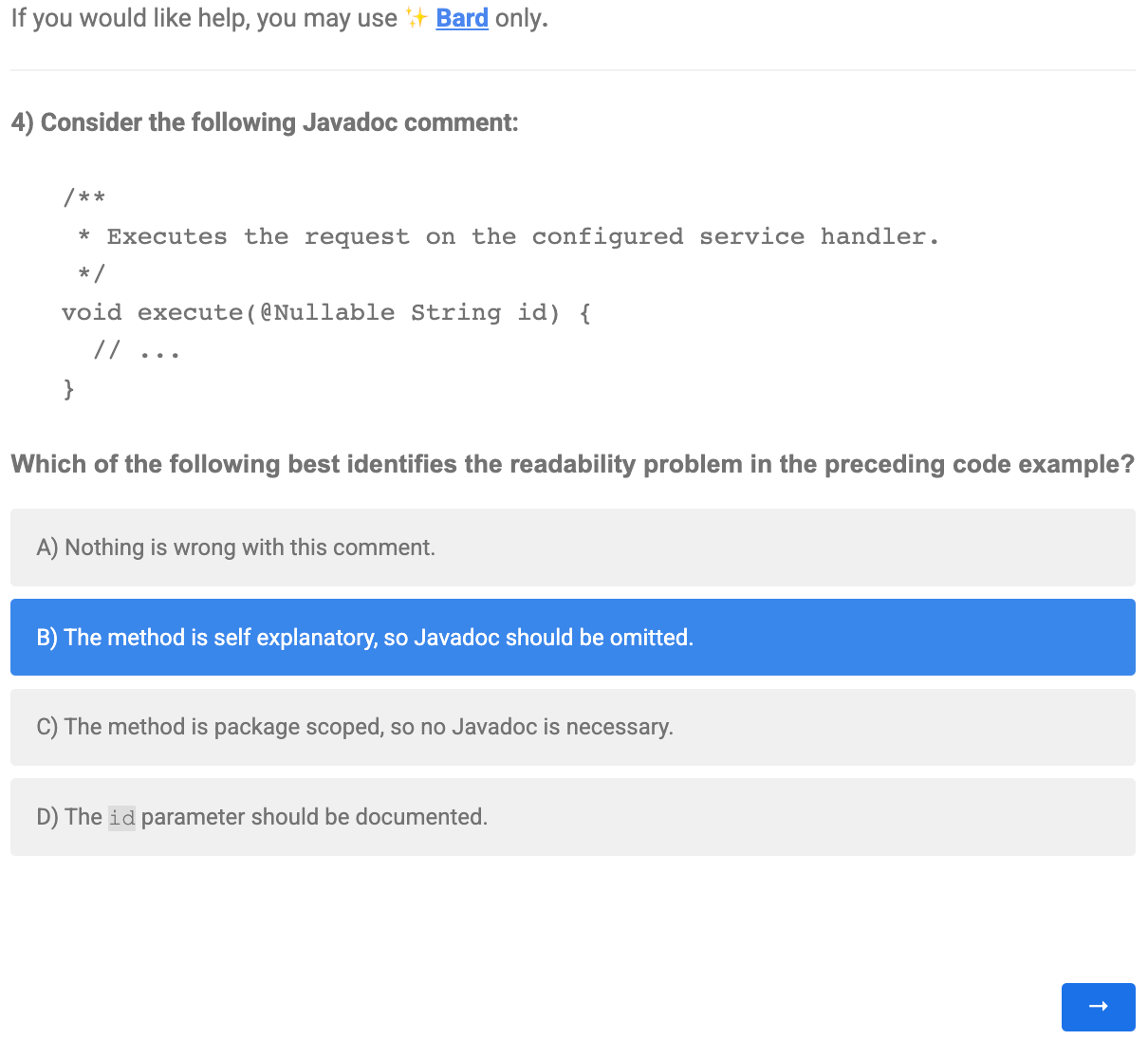}
        \caption{First pass on a Bard-first, solve-type question.}
        \label{fig:bard-first}
        \Description[A screenshot of the first pass on a Bard-first question on the Qualtrics exam website.]{A screenshot of the first pass on a Bard-first question on the Qualtrics exam website. The top of the screen says, "If you would like help, you may use Bard only. There is a link to the Bard webpage. In the middle of the screen is an exam question: "Consider the following Javadoc comment... [Code sample]... Which of the following best identifies the readability problem in the preceding code sample?" There are 4 multiple choice answers below the question, with "B) The methods are self-explanatory so a Javadoc is not required." selected and highlighted. There is a "Next" arrow at the bottom of the screen.}
    \vspace*{-6pt}\end{figure}

\begin{figure}
        \centering
        \includegraphics[width=.45\textwidth]{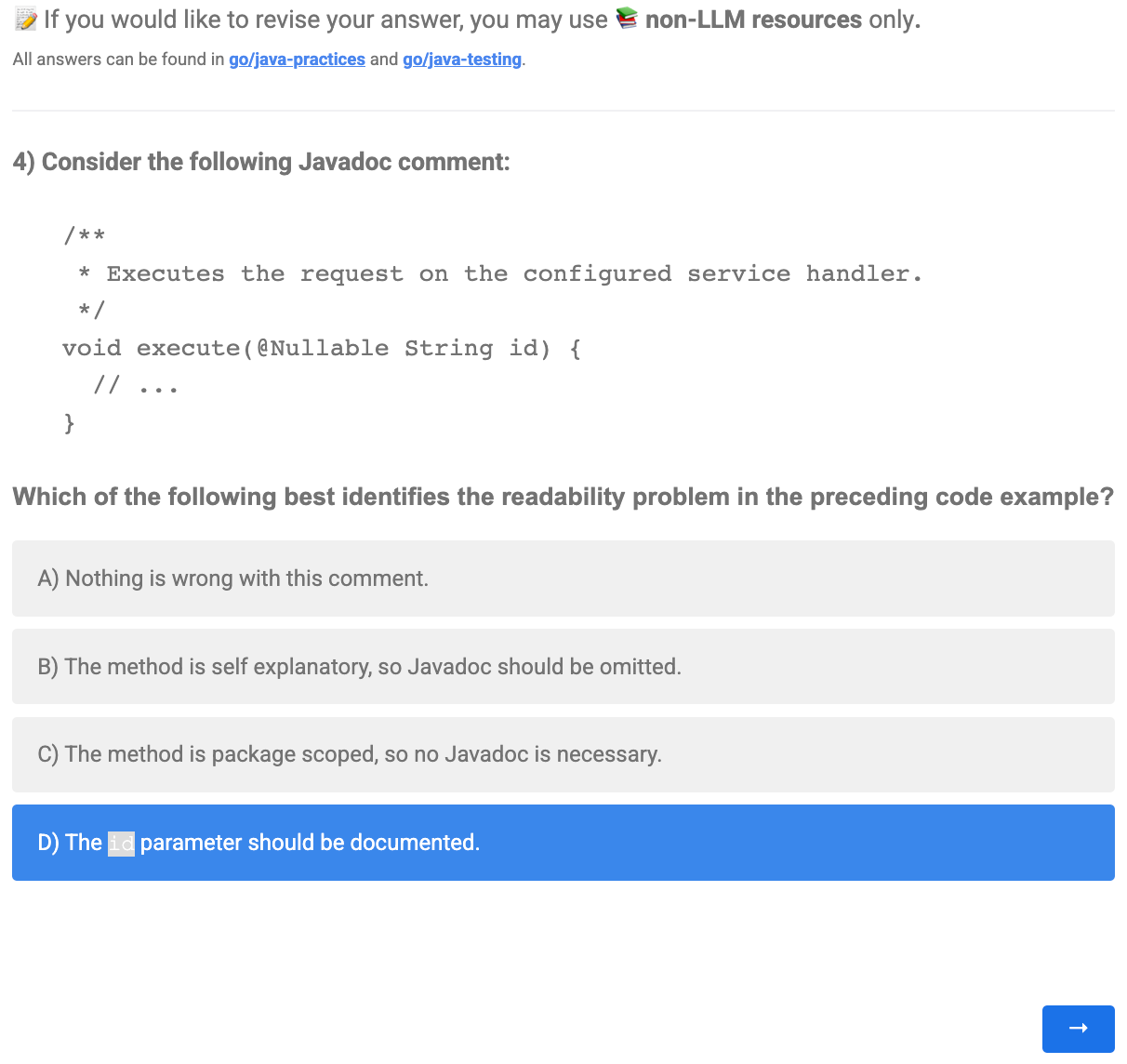}
        \caption{Second pass on a Bard-first, solve-type question.}
        \label{fig:book-next}
        \Description[A screenshot of the second pass on a Bard-first question on the Qualtrics exam website.]{A screenshot of the second pass on a Bard-first question on the Qualtrics exam website. The content is the same as before, with two notable differences. First, instead of "If you would like help, you may use Bard only", the header text reads, "If you would like to revise your answer, you may use non-LLM resources only." A link to the documentation is provided. Second, instead of B) being the selected multiple-choice answer, "D) The id parameter should be documented." is now selected, suggesting that the user changed their answer.}
   \vspace*{-6pt}\end{figure}

\textbf{Question types (search vs. solve):} There are five "search"-type and five "solve"-type questions; participants are not explicitly told these classifications. Search-type questions are more straightforward and have answers that map directly to the style guide. For example, the correct answer to \textit{“When would you declare a nested class as static?”} can be found near-verbatim in the chapter of the documentation about the "static" keyword. Solve-type questions are more open-ended and involve the critique of a provided code snippet; the answer cannot be found directly in the documentation. An example of a solve-type question is shown in \textit{Fig.s \ref{fig:bard-first}} and \textit{\ref{fig:book-next}}.

\subsection{Expertise measurement}\label{epxertise}
To evaluate variation in outcomes by expertise and perceived \nobreak expertise, we constructed both an objective and self-reported \nobreak expertise percentile rank within our sample. The objective expertise rank considers company-internal statistics such as amount of code written, tenure, Java experience, and previous "readability" exam experience. The perceived expertise rank considers participants’ self-assessment of previous experience (Java, LLM-based tools, productivity) measured during the pre-task survey. \textit{Appendix \ref{appendix:reexp}} shows the data and calculations for these ranks. We refer to \textit{experts} relative to \textit{novices} as those with a higher objective expertise percentile rank. 

\subsection{Thematic analysis}\label{analysis}
In addition to evaluating descriptive data from the screening/pre-task surveys and quantitative data from the exam using statistical methods, we conducted a thematic analysis \cite{Braun2006} on transcriptions of in-session think-aloud commentary and post-task survey responses. Initial codes for this analysis were generated from prior research, our 8 pilot studies, and the first batch of participant data. Themes about participant behaviors in this paper are saturated by codes that appeared in at least ten distinct user sessions \cite{Ando2014, Guest2006}. Representative quotes from this analysis supplement our findings.

\section{Results}\label{results}
\subsection{Productivity}

\begin{figure}
    \includegraphics[width=.5\textwidth]{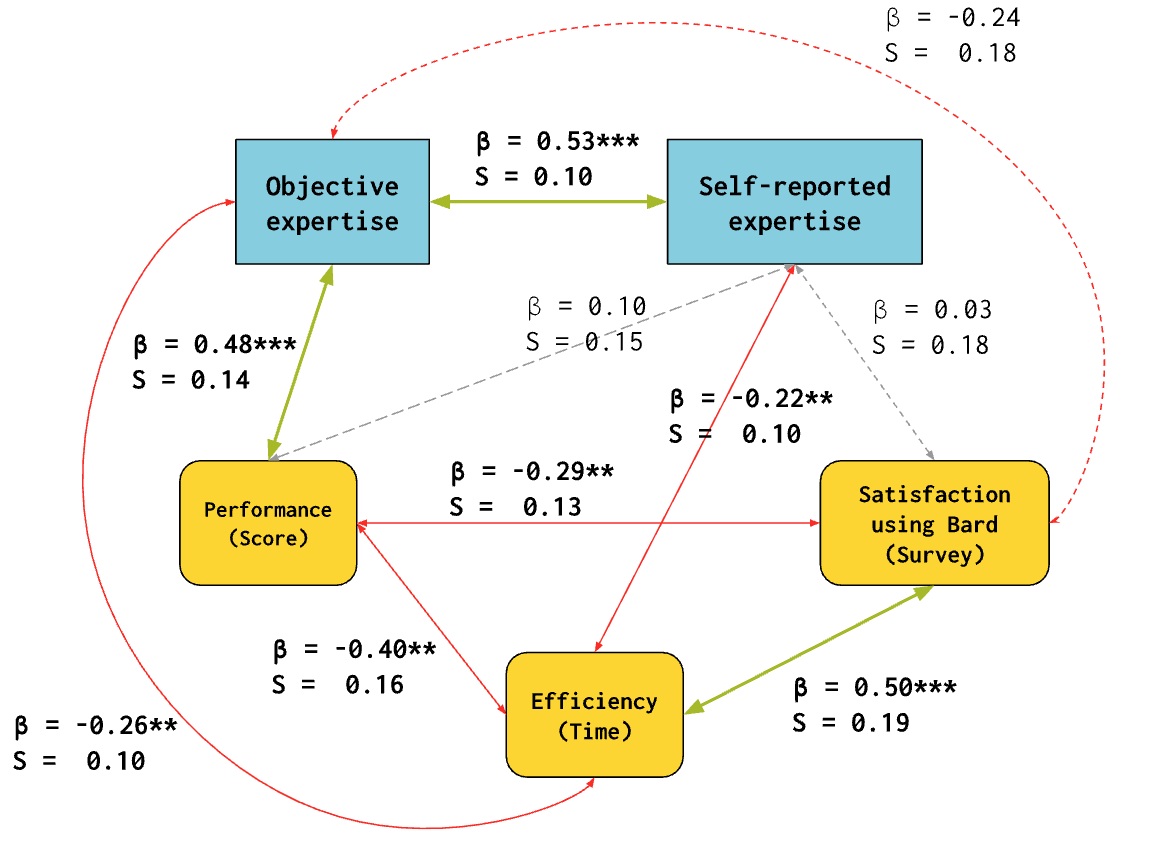}
    \caption{A structural equation model showing correlations between expertise measures and prodcutivity outcomes; $\beta$ is the normalized effect size in standard deviations, and $S$ denotes standard error.$^4$}
    \label{fig:sem}
    \Description[A structural equation model graph with five nodes.]{The graph shows five nodes: two expertise measures (objective and self-reported expertise) and three productivity measures (performance, efficiency, satisfaction). Each node is connected to every other node with a bi-directional line. Each line has a corresponding effect size, standard error value, and significance value indicating the direction and significance of the relationship. For example, objective expertise has a positive relationship with performance on the exam with 99\% confidence. However, there is no significant relationship between self-reported expertise and performance on the exam.}
\end{figure}

In \textit{Fig. \ref{fig:sem}}, we show variation in productivity outcomes by expertise through a structural equation model (SEM) \cite{Ullman2003} relating our two percentile rank-transformed expertise measures (objective and self-reported) and three productivity constructs. Moving forward in the paper, we use the \textit{objective} expertise percentile rank as the primary measurement of expertise in our discussion, as self-reported measures have low correlation with the outcomes of interest. Latent measures of productivity were constructed \cite{Forsgren2021} as follows:
 
 \begin{itemize}
    \item \textbf{Performance}: Total exam score (0–10). One point per \nobreak correct answer following the second pass. No partial credit.
    \item \textbf{Efficiency}: Total time spent on the exam.
    \item \textbf{Satisfaction}: Summed satisfaction score \\ aggregated from the post-task survey (\textit{Table \ref{table:sat}}).
\end{itemize}
%

\begin{figure}
         \centering
        \includegraphics[width=\columnwidth]{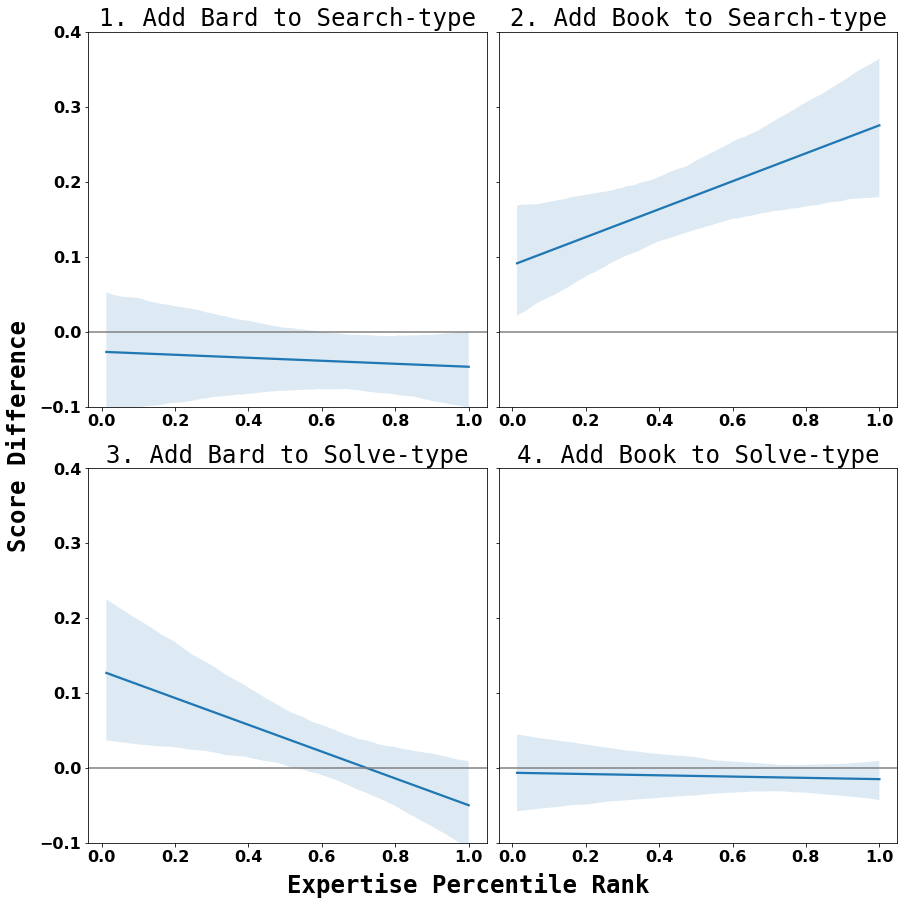}
            \captionof{figure}{An OLS regression of the difference in scores between passes on expertise, with 95\% confidence intervals.}
            \Description[A panel of four line graphs.]{A panel of four line graphs, labeled 1. Add Bard to Search-type, 2. Add Book to Search-type, 3. Add Bard to Solve-type, and 4. Add Book to Solve-type. On the x-axis is "Expertise Percentile Rank," with a range between 0.0 and 1.0. On the y-axis is "Score Difference", shown between -.1 and 0.4. Each line also has a confidence band. The only significant relationships are 2. Add Book to Search Type, where the score difference is positively increasing with expertise, and 3. Add Bard to Solve-type, where the score difference is negatively decreasing with expertise.}
        \label{fig:ols-graph}
    \end{figure}

\setcounter{footnote}{4}\footnotetext{Across all tables and diagrams, we use the following significance notation: *: p<.10, **: p<.05, ***: p<.01.}\subsubsection{Performance}\label{performance}

On average, participants scored 4.89 out of 10 points ($\sigma = 1.7$ points). There was no significant change in scores over time or between the passes of a question. Participants scored significantly higher on search-type questions. Measured expertise and final exam score have a significant positive relationship (\textit{Fig. \ref{fig:sem}}). 

Next, we calculate the effect of adding access to a resource across expertise by regressing the score difference between passes on expertise, adding fixed effects for treatment order (Bard-First vs. Bard-Last) and question type (search vs. solve). For example, a score difference of -1 indicates that the user changed from a correct to an incorrect answer between passes. These effect sizes are visualized in \textit{Fig. \ref{fig:ols-graph}} and described below.

\begin{enumerate}
    \item[(1)]\textbf{Adding Bard to a search-type question} had no \nobreak significant effect. Participants reported feeling confident in the answers they found using Book resources in the first pass, and would often skip their second pass with Bard.
\end{enumerate}
       \begin{quote}
    \textit{``Sometimes [the docs] covered the exact topic.''}\hspace{1em plus 1fill}---P12
        \end{quote}
    \begin{quote}
    \textit{``I trusted answers from the documentations more.. they were often more concrete\ldots for example, 'make all nested classes static.' When I found answers in the documentation, I was much [more confident] in them.''}\hspace{1em plus 1fill}---P77
        \end{quote}
\begin{enumerate}
    \item[(2)]\textbf{Adding Book to a search-type question} significantly \nobreak improved the score, especially for experts. Experts demonstrated more familiarity with navigating and finding relevant sections of documentation:
\end{enumerate}
        \begin{quote}
            \textit{``I know we have documentation on this.. I've used it before\ldots''}\hspace{1em plus 1fill}---P62
        \end{quote}
        \begin{quote}
            \textit{``Documentation was faster.. especially when I already had an idea of what the right answer was. 
    ''}\hspace{1em plus 1fill}---P64
        \end{quote}
         
    Novices demonstrated less familiarity and more difficulty in identifying and interpreting relevant documentation:
        \begin{quote}
            \textit{``Because I don't use Java, none of the [documentation] means much to me\ldots''}\hspace{1em plus 1fill}---P75
        \end{quote}
        \begin{quote}
            \textit{``I searched through the docs for what I lacked knowledge on, but \ldots I wasn't sure what to search for.''}\hspace{1em plus 1fill}---P63
        \end{quote}

\begin{enumerate}
\item[(3)]\textbf{Adding Bard to a solve-type question} improved \nobreak performance for novices, but had no effect for experts. The Book did not contain answers to solve-type questions, so \nobreak participants were largely reliant on their expertise. Experts were more likely to answer correctly in the first pass, benefiting less from Bard in the second pass.
      \item[(4)]\textbf{Adding Book to a solve-type question} had no \nobreak significant effect. Book resources had less specific guidance for critiquing a code sample:
\end{enumerate}
        \begin{quote}
            \textit{``This is the kind of task that Bard does really well on… I don’t know how to look in the docs for this.''}\hspace{1em plus 1fill}---P24
        \end{quote}
        \begin{quote}
            \textit{``I don't think the documentation would be useful for this case.''}\hspace{1em plus 1fill}---P69
        \end{quote}
        \begin{quote}
            \textit{``How would I even search for this?''}\hspace{1em plus 1fill}---P71
        \end{quote}

\subsubsection{Efficiency}\label{efficiency}

\begin{table*}
\begin{tabular}{cc}
    \includegraphics[width=.95\textwidth]{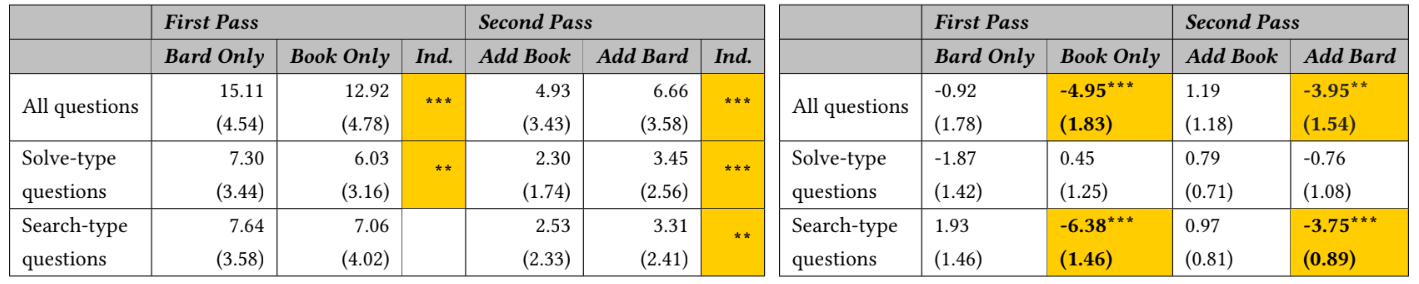}
\end{tabular}
    \captionof{table}{\textit{Left, a}) Descriptive statistics on efficiency. Mean time spent per section in minutes, with sd. in parentheses. The \textit{\textbf{Ind.}} column displays significance values from a two-sample independence t-test between Bard and Book times. \textit{Right, b}) Regression coefficients and standard errors from a regression of time spent (in minutes) on expertise percentile rank.$^5$}
    \label{table:time}
\end{table*}


On average, participants completed the exam in 39.6 minutes ($\sigma = 9$). Controlling for question number, there was a slight speedup per question of ~20 seconds. There’s no significant correlation between time per question and accuracy, but more time spent on the entire exam correlates with lower final score (\textit{Fig. \ref{fig:sem}}).

\textbf{Participants spent more time using Bard:}  Participants spent significantly more time using Bard, both in the first pass (Bard only) and in the second pass (adding Bard). This difference is more significant for solve-type questions (\textit{Table \ref{table:time}a}). Here are some factors as to why participants spent more time with Bard:\setcounter{footnote}{5}\footnotetext{These regressions are performed at the per-question level with relevant controls as stated (N=190 per question type, N=380 across all questions). Data appears normally distributed, and a t-test for independence is sufficient at this sample size \cite{fagerland2012}.}
\begin{itemize}
    \item \textbf{Slower response times:} Participants noticed and expressed frustration at more latency in AI response times compared to search query response times. This may be a transient issue as generative AI technology matures.
    \item \textbf{Less specific visual direction:} The documentation had clear headers that led the eye to the correct answer; \nobreak participants would stop reading after they identified the appropriate passage for a search-type question. Bard's \nobreak generated outputs were paragraph-like and verbose; the \nobreak correct guidance was more obscured within the text.
    \item \textbf{Verbose interactions:} Participants wrote in longer sentences and had more interactions with Bard following their initial query, in contrast to having brief keyword interactions when using other resources. For example, P77 queried for \textit{"IllegalArgumentException"} and P78 queried for \textit{“best practice autovalue java”} in a search engine. When they used Bard to answer the same questions, they wrote the following:
\end{itemize}
       \begin{quote}
            \textit{``Heya, could you please write me a Java function to assert that removing an item from a list throws an IllegalArgumentException? Thanks.''}\hspace{1em plus 1fill}---P77
        \end{quote}
      
        \begin{quote}\textit{``Give me code examples to show the best way to test for an expected exception\ldots Do so in a unit test \ldots I don't feel like this is correct?''}\hspace{1em plus 1fill}---P78, in a back-and-forth conversation\end{quote}

\textbf{Experts and novices spend similar time using Bard:} \textit{Table \ref{table:time}b} regresses time spent per question on expertise. When experts have access to Book resources in the first pass for a search-type question, they spend less time than novices on both passes. \nobreak However, when experts have access to Bard first, they do not interact with or interpret Bard outputs any faster than novices.

\textbf{Participants felt more efficient using Bard: } Despite
 spending more time using Bard, participants significantly agreed with the following in their post-survey assessment (\textit{Table \ref{table:sat}}): \textit{2. I complete tasks faster when using Bard, 3. I spend less mental effort when using Bard}, and \textit{4. I spend less time searching for information or examples when using Bard}.

\begin{quote}\textit{``For someone who knows very little about Java, Bard would speed up my workflow a lot. I would have to read a lot of the style guide and Bard makes things much faster.''}\hspace{1em plus 1fill}---P11\end{quote}

\begin{quote}\textit{``I definitely found [Bard] easier than searching the documentation \ldots  I found that using Bard was surprisingly effective... It seemed faster to use Bard because I could ask it more stream of consciousness questions.''}\hspace{1em plus 1fill}---P50\end{quote}

\subsubsection{Satisfaction}\label{satisfaction}
Participants felt significantly more productive when using Bard compared to Book  (\textit{Table \ref{table:sat}, Statement 1}). Novices agreed with this more so than experts,  perhaps because experts are more likely to identify flaws with the AI \cite{Ou2023}. However, there is no significant change in satisfaction (\textit{Statement 5}) or frustration (\textit{Statement 6}) following the task.\setcounter{footnote}{6}\footnotetext{Mean values $\mu$ indicate the degree of agreement with the statement, calculated using the method described in \textit{Appendix \ref{appendix:jian}}. Significance values are calculated from a one-sample t-test with the null hypothesis that $\mu = 0$, and regression coefficients $\beta$ demonstrate the relationship between response values and expertise. Highlighted cells indicate significant findings.}\setcounter{footnote}{6}
\aptLtoX[graphic=no,type=html]{\begin{table}[H]
        \centering
    \begin{tabular}{|l|l|l|}
    \hline
    \rowcolor[HTML]{C0C0C0} 
    \textbf{Compare how you felt} \break \textbf{when completing tasks with Bard assistance,} \break \textbf{as opposed to without Bard assistance.} &      $\mu$ &      $\beta$ \\ \hline
    1. I am more productive when using Bard. &
      \cellcolor[HTML]{FFCD00}{\color[HTML]{212121} \textbf{.24**}\break \textbf{(0.92)}} &
      \cellcolor[HTML]{FFCD00}{\color[HTML]{212121} \textbf{-0.77*}\break \textbf{(0.40)}} \\ \hline
    2. I complete tasks faster when using Bard. &
      \cellcolor[HTML]{FFCD00}{\color[HTML]{212121}  \textbf{.04***}\break \textbf{(1.11)}} &
       -0.66\break (0.48)  \\ \hline
    3. I spend less mental effort when using Bard. &
      \cellcolor[HTML]{FFCD00}{\color[HTML]{212121} \textbf{0.46***}\break \textbf{(1.19) }} &
      -0.03\break (0.49) \\ \hline
    4. I spend less time searching for information\break    or examples when using Bard. &
      \cellcolor[HTML]{FFCD00}{\color[HTML]{212121} \textbf{0.51***}\break \textbf{(1.06)}} &
      -0.35\break (0.44) \\ \hline
    5. I feel more satisfied with my work completing\break     this task when using Bard. &
      -0.11\break (0.96) &
      -0.57\break (0.41) \\ \hline
    6. I find myself less frustrated when using Bard. &
       0.07\break (0.99)  &
       0.26\break (0.43)  \\ \hline
    \end{tabular}%
    \caption{Comparative satisfaction survey \cite{Ziegler2022} administered post-task.$^6$}
    \label{table:sat}
  \end{table}}
{\begin{table}[H]
        \centering
        \resizebox{\columnwidth}{!}{%
    \begin{tabular}{|l|l|l|}
    \hline
    \rowcolor[HTML]{C0C0C0} 
    \begin{tabular}[c]{@{}l@{}}\textbf{Compare how you felt} \\ \textbf{when completing tasks with Bard assistance,} \\ \textbf{as opposed to without Bard assistance.}\end{tabular} &
      $\mu$ &
      $\beta$ \\ \hline
    1. I am more productive when using Bard. &
      \cellcolor[HTML]{FFCD00}{\color[HTML]{212121} \begin{tabular}[c]{@{}l@{}}\textbf{.24**}\\ \textbf{(0.92)}\end{tabular}} &
      \cellcolor[HTML]{FFCD00}{\color[HTML]{212121} \begin{tabular}[c]{@{}l@{}}\textbf{-0.77*}\\ \textbf{(0.40)}\end{tabular}} \\ \hline
    2. I complete tasks faster when using Bard. &
      \cellcolor[HTML]{FFCD00}{\color[HTML]{212121} \begin{tabular}[c]{@{}l@{}}\textbf{.04***}\\ \textbf{(1.11)}\end{tabular}} &
      \begin{tabular}[c]{@{}l@{}}-0.66\\ (0.48)\end{tabular} \\ \hline
    3. I spend less mental effort when using Bard. &
      \cellcolor[HTML]{FFCD00}{\color[HTML]{212121} \begin{tabular}[c]{@{}l@{}}\textbf{0.46***}\\ \textbf{(1.19)}\end{tabular}} &
      \begin{tabular}[c]{@{}l@{}}-0.03\\ (0.49)\end{tabular} \\ \hline
    \begin{tabular}[c]{@{}l@{}}4. I spend less time searching for information\\     or examples when using Bard.\end{tabular} &
      \cellcolor[HTML]{FFCD00}{\color[HTML]{212121} \begin{tabular}[c]{@{}l@{}}\textbf{0.51***}\\ \textbf{(1.06)}\end{tabular}} &
      \begin{tabular}[c]{@{}l@{}}-0.35\\ (0.44)\end{tabular} \\ \hline
    \begin{tabular}[c]{@{}l@{}}5. I feel more satisfied with my work completing\\     this task when using Bard.\end{tabular} &
      \begin{tabular}[c]{@{}l@{}}-0.11\\ (0.96)\end{tabular} &
      \begin{tabular}[c]{@{}l@{}}-0.57\\ (0.41)\end{tabular} \\ \hline
    6. I find myself less frustrated when using Bard. &
      \begin{tabular}[c]{@{}l@{}}0.07\\ (0.99)\end{tabular} &
      \begin{tabular}[c]{@{}l@{}}0.26\\ (0.43)\end{tabular} \\ \hline
    \end{tabular}%
    }
    \caption{Comparative satisfaction survey \cite{Ziegler2022} administered post-task.$^6$}
    \label{table:sat}
  \end{table}
}

\subsection{Trust}

We measure trust using both demonstrated measures (actions taken) and perceived measures (self-assessment) \cite{Kohn2021}. 

\subsubsection{Demonstrated measures of trust}\label{dem-trust} Users trust and depend upon a resource when they delegate and rely on it \cite{Xie2019, Lemmer2023} and distrust when they reject it \cite{Para1997}.\footnote[7]{With respect to our paper title, participants can use the AI (\textit{Take it}), reject the AI (\textit{Leave it}), or change their answer following usage of the AI (\textit{Fix it}).}\setcounter{footnote}{7} These actions in our study are described in \textit{Table \ref{table:actions}}. Participants can trust a resource either correctly or incorrectly;\footnote{Incorrectly trusting is also referred to as \textit{mistrust} or \textit{overtrust} in literature \cite{Lee2004}.} this attribution depends on the resulting score. 

\begin{figure} 
    \centering
    \includegraphics[width=\columnwidth]{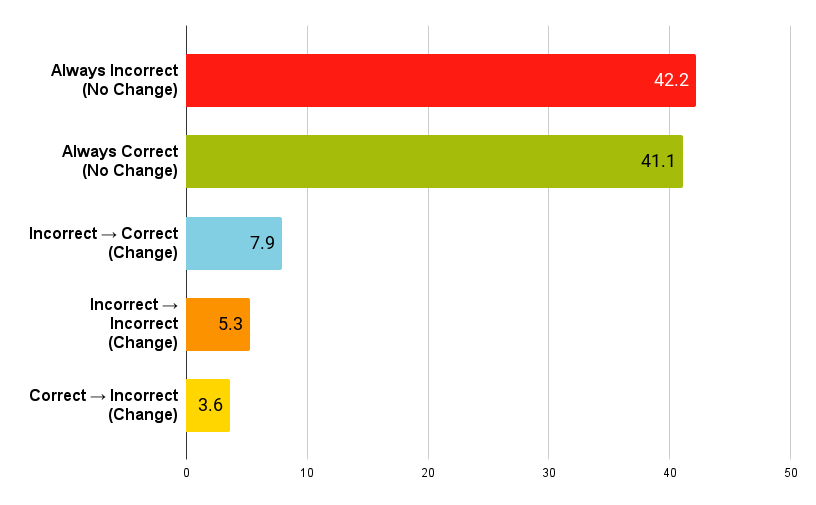}
    \caption{Answer-changing percentages between passes.}
    \Description[A bar chart of the 5 possible actions and outcomes.]{A bar chart of the 5 possible actions and outcomes. In 42.2\% of the 760 questions, the answer was always incorrect (no-change). 41.1\% were always correct (no change). 7.9\% of the time, an incorrect answer was changed to a correct answer, and a correct answer was changed to an incorrect answer 3.6\% of the time. An incorrect answer was changed to another incorrect answer 5.3\% of the time.}
\label{fig:change-stats}
  \end{figure}

\textit{Fig. \ref{fig:change-stats}} shows the percentage of answer-changing behaviors \nobreak between passes; participants often do not change their answer across passes. When the correctness of a participant’s answer does not change between passes, the trust implication is ambiguous: for example, if a participant got the correct answer during both passes, they could have already known the correct answer, relied upon either or both resources, or used but dismissed those resources.\footnote{In our experimental design, we considered asking participants which of these scenarios were the case after each question, but this added considerable time to each user session. Pilot participants demonstrated fatigue after the hour-mark.}
\vspace{5pt}

   \begin{table}%
    \centering
\resizebox{\columnwidth}{!}{%
\begin{tabular}{|l|l|l|l|}
\hline
\cellcolor[HTML]{C0C0C0}\  &\cellcolor[HTML]{C0C0C0} \textbf{First pass} &\cellcolor[HTML]{C0C0C0} \textbf{Second pass} &\cellcolor[HTML]{C0C0C0} \textbf{Behavior/Implication}                                                                       \\ \hline
1 & Skip                & Skip                 & Distrusted Both                                                                            \\ \hline
2 & Skip                & Bard                 & Distrusted Book, trusted Bard                                                                            \\ \hline
3 & Skip                & Book                 & Distrusted Bard, trusted Book                                                                            \\ \hline
4 & Bard                & Skip                 & Distrusted Book, trusted Bard                                                                            \\ \hline
5 & Book                & Skip                 & Distrusted Bard, trusted Book                                                                            \\ \hline
6 & Bard                & Book                 & \begin{tabular}[c]{@{}l@{}}If the answer changed: trusted Book\end{tabular} \\ \hline
7 & Book                & Bard                 & \begin{tabular}[c]{@{}l@{}}If the answer changed: trusted Bard\end{tabular} \\ \hline
\end{tabular}%
}
    \caption{Action space and implications. A participant \textit{uses} a resource if they click into it during the pass. A participant \textit{skips} if they do not click into the resource during the pass.}
    \label{table:actions}
  \end{table}

\textbf{How does trust change over time?}
\textit{Table \ref{table:trust}a} shows a \nobreak regression of trust actions on question order, with expertise and question number fixed effects.\footnote{Tables 4 and 5 show regression coefficients with standard errors in parentheses.} Experts significantly depended on Book resources more so than novices, and dependence on Book resources did not change over time. All participants, particularly novices, increased dependence on Bard over the course of the exam, despite reporting that they wanted to decrease Bard usage:

\begin{quote}\textit{``So I got it right, and then I got it wrong with Bard.. maybe I shouldn't trust Bard then.
''}\hspace{1em plus 1fill}---P13\end{quote}
\begin{quote}\textit{``I don't think what Bard is saying is true... I'll probably stop using Bard, just because it's incorrect.''}\hspace{1em plus 1fill}---P35\end{quote}

\begin{table*}
\begin{tabular}{cc}
    \includegraphics[width=.95\textwidth]{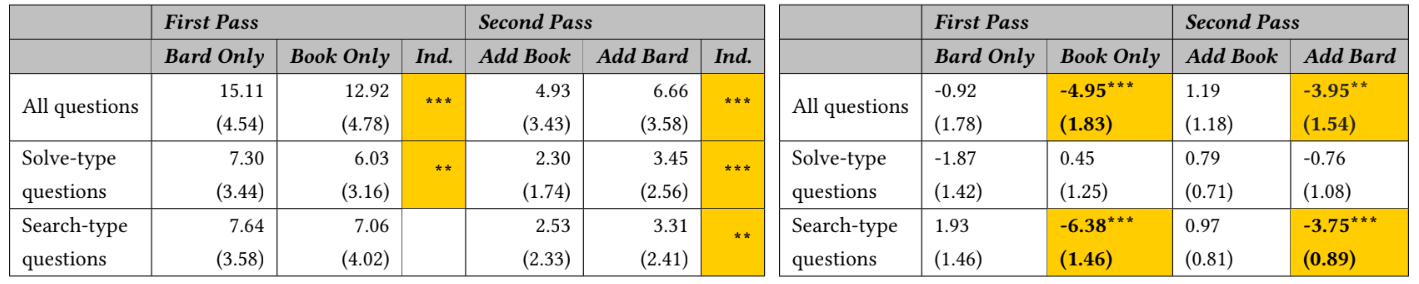}
\end{tabular}
    \caption{\textit{Left, a}) Regression of trust actions on question order (over time). \textit{Right, b)} Regression of the likelihood of taking a correctly trusting or incorrectly trusting action (\textit{Table \ref{table:actions}}) on expertise.}  
    \label{table:trust}
\end{table*}

\textbf{Who correctly and incorrectly trusts?} \textit{Table \ref{table:trust}b} regresses the likelihood of taking a correctly trusting or incorrectly trusting action on expertise. Participants of all expertise are equally likely to be led astray and incorrectly trust Bard. Novices are slightly more likely to correctly trust Bard across both types of questions.

\begin{table}\tabcolsep4pt
\resizebox{\columnwidth}{!}{%
\begin{tabular}{|lllp{60pt}|}
\hline
\multicolumn{1}{|c|}{\cellcolor[HTML]{C0C0C0} \ \ ~ } &
  \multicolumn{1}{>{\centering}p{70pt}|}{\cellcolor[HTML]{C0C0C0}\textbf{Distrusted both}\newline \cellcolor[HTML]{C0C0C0}\textbf{(Self-trust)}} &
  \multicolumn{1}{p{60pt}|}{\cellcolor[HTML]{C0C0C0}\textbf{Distrusted}\newline \cellcolor[HTML]{C0C0C0}\textbf{Bard}} &
  \multicolumn{1}{p{60pt}|}{\cellcolor[HTML]{C0C0C0}\textbf{Distrusted}\newline \cellcolor[HTML]{C0C0C0}\textbf{Book}} \cr \hline
\multicolumn{4}{|l|}{\cellcolor[HTML]{EFEFEF}\textit{1. Effect of expertise on the likelihood of expressing distrust}} \cr \hline
\multicolumn{1}{|l|}{All questions} &
  \multicolumn{1}{p{70pt}|}{\cellcolor[HTML]{FFCD00}\textbf{0.12***}\newline \cellcolor[HTML]{FFCD00}\textbf{(0.04)}} &
  \multicolumn{1}{p{60pt}|}{\cellcolor[HTML]{FFCD00}\textbf{0.10***}\newline \cellcolor[HTML]{FFCD00}\textbf{(0.04)}} &
  \cellcolor[HTML]{FFCD00}\textbf{-0.20***}\newline \cellcolor[HTML]{FFCD00}\textbf{(0.05)} \cr \hline
\multicolumn{1}{|l|}{Solve-type questions} &
  \multicolumn{1}{p{70pt}|}{\cellcolor[HTML]{FFCD00}\textbf{0.16***}\newline \cellcolor[HTML]{FFCD00}\textbf{(0.06)}} &
  \multicolumn{1}{p{60pt}|}{0.03\newline (0.04)} &
  \cellcolor[HTML]{FFCD00}\textbf{-0.24***}\newline \cellcolor[HTML]{FFCD00}\textbf{(0.08)} \cr \hline
\multicolumn{1}{|l|}{Search-type questions} &
  \multicolumn{1}{p{70pt}|}{\cellcolor[HTML]{FFCD00}\textbf{0.09*}\newline\cellcolor[HTML]{FFCD00}\textbf{(0.5)}} &
  \multicolumn{1}{p{60pt}|}{\cellcolor[HTML]{FFCD00}\textbf{0.17***}\newline \cellcolor[HTML]{FFCD00}\textbf{(0.06)}} &
  \cellcolor[HTML]{FFCD00}\textbf{-0.17**}\newline \cellcolor[HTML]{FFCD00}\textbf{(0.07)} \cr \hline
\multicolumn{4}{|l|}{\cellcolor[HTML]{EFEFEF}\textit{2. Effect of expressing distrust on performance}} \cr \hline
\multicolumn{1}{|l|}{All questions} &
  \multicolumn{1}{p{70pt}|}{0.01\newline (0.06)} &
  \multicolumn{1}{p{60pt}|}{\cellcolor[HTML]{FFCD00}\textbf{0.24***}\newline \cellcolor[HTML]{FFCD00}\textbf{(0.06)}} &
  \cellcolor[HTML]{FFCD00}\textbf{-0.17***}\newline  \cellcolor[HTML]{FFCD00}\textbf{(0.04)} \cr \hline
\multicolumn{1}{|l|}{Solve-type questions} &
  \multicolumn{1}{p{70pt}|}{0.09\newline (0.08)} &
  \multicolumn{1}{p{60pt}|}{0.11\newline (0.11)} &
  -0.01\newline (0.06) \cr \hline
\multicolumn{1}{|l|}{Search-type questions} &
  \multicolumn{1}{p{70pt}|}{-0.12\newline (0.08)} &
  \multicolumn{1}{p{60pt}|}{\cellcolor[HTML]{FFCD00}\textbf{0.16**}\newline \cellcolor[HTML]{FFCD00}\textbf{(0.07)}} &
  \cellcolor[HTML]{FFCD00}\textbf{-0.30***}\newline \cellcolor[HTML]{FFCD00}\textbf{(0.06)} \cr \hline
\end{tabular}%
}
    \caption{Regression coefficients on distrust behavior.}
\label{table:distrust}
\end{table}

\textbf{Who exhibits distrust?} \textit{Table \ref{table:distrust}.1} regresses the likelihood of expressing distrust on expertise, conditional on question effects (question number and question order). As expertise increases, the likelihood of distrusting both resources and distrusting Bard increases, and the likelihood of distrusting Book resources decreases. That is, novices were more likely than experts to distrust Book resources, and experts were more likely to distrust Bard. 

\textbf{Is it a good strategy to distrust?} \textit{Table \ref{table:distrust}.2}
regresses score per question on likelihood of distrusting, conditional on question effects and expertise. For solve-type questions, distrusting either resource had no effect on the score. For search-type questions where the answer could be found in the Book, relying on the Book yielded better scores. Better performance by experts may be partially attributed to experts knowing when to appropriately distrust Bard and rely on the Book for Search-type questions. However, this distrust behavior could also punish performance; as expertise increased, the likelihood of using Bard to correct an incorrect answer decreased.

\subsubsection{Perceived measures of trust}\label{survey-trust}
We administered Jian's Trust in Automated Systems survey \cite{Jian2000} both before and after the task. Survey questions and responses are shown in \textit{Appendix \ref{appendix:jian}}. 
\par
\textbf{Participants calibrated trust post-task:} Before the task, \nobreak participants' sentiments towards Bard were mostly neutral. \nobreak Following the task, participants expressed significantly less trusting sentiments towards Bard. This change is likely due to participants \textit{appropriately calibrating trust} rather than \textit{losing trust}, given that users tend to replace dispositional, pre-exposure trust in automation following exposure to that system with feedback \cite{Merritt2008, Feng2023}.


Most of our sample did not have significant prior experience interacting with the AI; $60.7\%$ reported in the pre-survey that they had rarely or never used LLM-assisted tools in development tasks at work, and many reiterated this during the task.\footnote{This may be due to a company-internal policy to avoid input of business-sensitive information and code into conversational AI agents.}

\section{Discussion}\label{discussion}

In this section, we discuss the key insights from our results, supported by literature review and patterns from our thematic analysis. Sections \ref{discuss:1} through \ref{discuss:3} touch on \textbf{RQ1: Effects on productivity}, digging into why we observed mixed results across different dimensions of productivity. Sections \ref{discuss:4} and \ref{discuss:5} touch on \textbf{RQ2: Behaviors of trust}, offering more context on participants' decision making behavior. Finally, in \ref{discuss:6}, we synthesize our observations into design implications, intended to aid developers in improving the design of intelligent conversational systems.

\subsection{Why might using the AI hurt performance?}\label{discuss:1} If users were rational, processed information optimally, and had perfect information, receiving more information would be better than receiving less information \cite{Simon1986}. However, receiving more \nobreak information at times hurt performance during the task: adding Bard did not strictly improve score (\textit{Fig. \ref{fig:ols-graph}}) and participants would at times change their answer from a correct to an incorrect answer after consulting additional resources (\textit{Fig. \ref{fig:change-stats}}). This behavior \nobreak penalized experts more, as experts were likely to get the correct \nobreak answer in the first pass. Here are some commonly observed behaviors that may indicate why participants might switch from a correct to an incorrect answer following a consultation with the AI:
\begin{enumerate}
    \item[(1)]\textbf{Users may not perceive the downsides of eliciting a second opinion.} Experts and novices were equally susceptible to changing an answer from correct to incorrect. Because this task was objectively scored based on specific correct answers, adding resources after already deriving the correct answer could only lead participants astray, not make them more correct. Participants did not seem to realize this pitfall:
\end{enumerate}
     \begin{quote}
            \textit{``I can use Bard, since it's available.. why not?''}\hspace{1em plus 1fill}---P66
    \end{quote}
    \begin{enumerate}
    \item[(2)]\textbf{Participants exhibit confirmation bias.}
Because participants could not copy multiple-choice answers easily due to infrastructure limitations\footnote{Often, participants' first instinct was to attempt to copy all answers into Bard. Several participants expressed frustration upon discovering that copying answers was disabled in the Qualtrics platform. From P21: \textit{I can't copy the answers, so it would be too much effort to ask [Bard].} Four participants even inspected the elements within the web browser to copy-paste the source code as a workaround. One participant painstakingly typed out each multiple-choice answer for each question, but abandoned this strategy due to time constraints.}, they would instead ask pointed questions to Bard such as\textit{ “Is A) \ldots the right answer?” } \nobreak Participants would seek out agreement and end their line of inquiry after receiving an affirming response, consistent with behavior exhibited in similar studies \cite{Cau2023, Mozannar2023}.
\end{enumerate}
\begin{quote}
            \textit{``Let's go with Bard, [because] this time, Bard and I agree on the answer.''}\hspace{1em plus 1fill}---P37
        \end{quote}
        \begin{quote}
            \textit{``[My strategy was that] I would ask Bard if they agreed that my answer was correct.''}\hspace{1em plus 1fill}---P28
        \end{quote}
    \vspace{2pt}
However, when participants prompted Bard without giving \nobreak sufficient context, Bard could make a case for affirming any of the provided answers.
\vspace{2pt}
\begin{quote}
            \textit{``[Bard] is not being very helpful because it's just \nobreak validating everything I'm saying.''}\hspace{1em plus 1fill}---P60
        \end{quote}
        \begin{quote}
            \textit{``In multiple cases, we had scenarios where [Bard] would confirm all of the answers\ldots''}\hspace{1em plus 1fill}---P12
        \end{quote}
        \vspace{2pt}
Participants found better success in asking comparative questions, e.g. \textit{“Which is the bigger problem, X or Y?”}
\vspace{2pt}
\begin{quote}
            \textit{``“[I'd] either ask Bard for free-form improvements to the code sample (not very reliable), or gave Bard the code and a couple answers I'm undecided between and ask Bard to pick between them (more reliable)\ldots''}\hspace{1em plus 1fill}---P60
        \end{quote}
        \begin{quote}
            \textit{``Maybe Bard is overeager to please- we got a yes on \nobreak every single one of these when asked individually. Maybe I'd get a more discriminating answer if I put in multiple options.''}\hspace{1em plus 1fill}---P12
        \end{quote}

\subsection{Why do participants perceive that they are more productive and efficient with the AI?}\label{discuss:2}

Despite the mixed measured effects on productivity, participants still perceived increased productivity and efficiency when using the AI (\textit{Table \ref{table:sat}}), perhaps due to reduced cognitive load.

\textbf{Using the AI is easy}: When using Book resources, participants would spend time \textit{actively}: searching for answers, skimming the text, and thinking about how to phrase keywords. When using Bard, participants spent time more \textit{passively}, waiting for responses and reading outputs \cite{Mozannar2023}. Participants across expertise levels exhibited effort substitution \cite{Xiao2023} by blindly copy-pasting questions.
\begin{quote}
    \textit{``I feel a little tired, so I will just start copy-pasting the question.''}\hspace{1em plus 1fill}---P19
\end{quote}
\begin{quote}
    \textit{``I would use Bard to reduce the cognitive load.''}\hspace{1em plus 1fill}---P31
\end{quote}
\begin{quote}
    \textit{``I liked Bard because you didn't have to do too much, you could just copy-paste and it would potentially find the answer that I'm looking for.''}\hspace{1em plus 1fill}---P47
\end{quote}
\begin{quote}
    \textit{``I guess I’ll actually read the question now while Bard is thinking.''}\hspace{1em plus 1fill}---P74
\end{quote}

\textbf{Users exhibit automation complacency:} If the AI produced optimal responses and users substituted effort appropriately, effort substitution would not be detrimental. However, given that we do not find strictly positive effects of access to the AI on productivity, this scenario meets Parasuraman's three requirements for automation complacency \cite{Para1997}\footnote{We cannot make a similar claim about \textit{automation bias}, which is defined as evidence of omission and commission errors when decision aids are imperfect \cite{Para2010}. We observe evidence of both errors; participants fail to notice omissions by the agent, and can be actively misled by the agent. However, this definition may be more appropriate for traditional automation; users have much more control over the outputs of conversational AI, so errors could be the result of either automation bias or imperfect usage.}: (1) A human operator is monitoring an automated system. (2) The frequency of such monitoring is lower than optimal. (3) There is a directly observable (negative) effect on performance.

\subsection{Why is there no change in satisfaction or frustration?}\label{discuss:3}

Despite feeling more productive and efficient, participants were not more satisfied or less frustrated when using the AI (\textit{Table \ref{table:sat}}), perhaps because negative emotional reactions and inappropriate sentiments of trust offset the perceived gains.

\textbf{Participants attribute blame asymmetrically:} When participants missed a question using Book resources, they were less likely to offer an explanation or attribute the fault to the resource directly. When they missed a question using Bard, they were more likely to display defensive behavior, justify efforts, and blame Bard or their ability to interpret and prompt Bard.
   \begin{quote}
        \textit{``Maybe I interpreted Bard's outputs wrong.''}\hspace{1em plus 1fill}---P29
    \end{quote}
   \begin{quote}
        \textit{``I'm not the master at prompting LLMs yet.''}\hspace{1em plus 1fill}---P47
    \end{quote}
   \begin{quote}
        \textit{``Bard is leading me astray! But I don't know how to tell if it's telling me stuff incorrectly.''}\hspace{1em plus 1fill}---P49
    \end{quote}

\textbf{Participants perceive Bard as a collaborator:} \textit{“Emotional \nobreak reactions may be a key element of trust and the decision to rely on automation \ldots”} \cite{Lee2004} and can be activated through perceived collaborations with automation \cite{Kuttal2021, Weisz2022}.
    \begin{quote}
        \textit{``Looks like Bard and I were wrong for this one.''}\hspace{1em plus 1fill}---P35
    \end{quote}
   \begin{quote}
        \textit{``[Bard] is like working with a pretty well-informed tutor. It's highlighting problems that are deeper than the multiple-choice questions and answers we were looking for.''}\hspace{1em plus 1fill}---P39
    \end{quote}
    \begin{quote}
        \textit{``We don't have many colleagues in the office these days\dots sometimes, it's much faster to ask colleagues since they have context. But given they're not there, I always go with Bard. ''}\hspace{1em plus 1fill}---P66
    \end{quote}

Further evidence of perceived collaboration was found as participants described Bard using language such as \textit{``It probably doesn't like me''} -P10 and \textit{``Bard is probably overwhelmed''} -P12. However, \textit{"using speech to create a conversational partner.. may lead people to attribute human characteristics to the automation in such a way as to induce false expectations that could lead to inappropriate trust."} \cite{Lee2004} This inappropriate trust, in turn, can induce less satisfaction with the AI \cite{Xiao2023}. 

\subsection{How do participants use resources?}\label{discuss:4} The overwhelmingly most common behavior was copy-pasting questions directly into Bard. Participants would also \textit{prime} the agent by adding context (e.g. \textit{“You are reviewing this code according to the [company] style guide”}). When using Book resources, participants would often search for keywords found in the question, either by manually skimming the documentation or querying external or company-internal search engines. If the question was \textit{“When would you declare a nested class as static?”}, participants might search for tokens such as \textit{“nested class static”}. They tended to not copy-paste a question directly into a search engine, especially when it was a solve-type question with code snippets.

\subsection{How do participants pick what to use?}\label{discuss:5}
The decision to use either resource was significantly correlated with question type and user expertise. For both search-type and solve-type questions, participants, particularly novices, preferred to consult the AI first. Participants reported wanting to use both resources concurrently, specifically by using Bard, then Book, then Bard as a sanity check.

\textbf{The AI reduces search frictions for novices:} Consulting the AI as a jumping off point helped novices to identify where to look in the documentation based on its recommendation.
        \begin{quote}
            \textit{``When I really did not have an idea\ldots Bard was helpful, because it looks broadly whereas searches had to be very precise.''}\hspace{1em plus 1fill}---P64
        \end{quote}
        \begin{quote}
            \textit{``I use [Bard] for more open-ended/general \nobreak questions\ldots When I have a specific question, I go to sources that are written and more concrete. I use Bard when there's something I don't know.''}\hspace{1em plus 1fill}---P60
        \end{quote}
        \begin{quote}
            \textit{``I would prefer to use Bard first, so I can ask a \nobreak general question to Bard. It's fast and gives me an answer quickly\ldots and if I am not happy with Bard's answer, I can do research on my own.''}\hspace{1em plus 1fill}---P72
        \end{quote}

\textbf{Users assume that the AI has limitations:} Many participants assumed that because Bard is trained on publicly available data, it would not be familiar with company-internal coding conventions. None checked to see whether the style guide was publicly available or tried to validate if Bard was familiar with the company’s Style Guide. Instead, they rejected the AI based on their assumption, perhaps because there is no way for users to concretely verify whether the AI is trained on any specific data source.

    \begin{quote}
        \textit{``I feel like Bard is an external tool, so for a task like readability [which is internal], it might not know the answers. For myself, if I want to find the answers, I'd just use [internal search], because I feel like external tools don't apply to our standards.''}\hspace{1em plus 1fill}---P6
    \end{quote}
    \begin{quote}
        \textit{``Because this is readability for [our company], and maybe what Bard gives me is general readability advice, so maybe it's not so useful for [our] readability questions\ldots''}\hspace{1em plus 1fill}---P78
    \end{quote}

\subsection{Design implications for more effective conversational AI}\label{discuss:6}
Despite Bard having the capability to perform better than the \nobreak average user on this task when questions and answers were directly copy-pasted as inputs, participants did not employ this strategy, perhaps due in part to the burden of increased effort exertion.\footnote{Some participants also suggested that having agency in the task felt important. From P47: \textit{``Copying things directly into Bard is a little silly.''}} Furthermore, the optimal strategy is dependent on both the context of the task and the expertise of the user relative to the capabilities of the agent. Feedback is given ex-post, which can make determining the optimal strategy intractable for the user. This suggests that there is room for improvement on behalf of the AI system to improve productivity; we recommend the following ideas for developers of conversational AI systems. 

\subsubsection{Design for appropriate trust.}\label{recs:1} Users appropriately trust \nobreak systems when they reject incorrect advice and accept correct advice \cite{Weisz2021, Sun2022}. Designing for \textit{appropriate} trust, not \textit{greater} trust \cite{Lee2004} can help mitigate to overreliance \cite{Itoh2011}.
\begin{enumerate}
    \item \textbf{Lower the degree of confidence.}
    Generative models can be perceived as overconfident, which can cause users of all expertise levels to display inappropriate trust. Communicating uncertainty on behalf of the agent can reduce undue overreliance \cite{Prabhudesai2023, Ross2023}. However, this may also force a user's cognitive effort, which can decrease satisfaction \cite{Bucinca2021}.
\end{enumerate}    
   \begin{quote}
        \textit{``I feel like [Bard] confused me\ldots it's so confident when it's wrong, so it's hard to follow your own barometer. The confidence scares me because it's just as confident when it's wrong compared to when it's right, and I'm bad at refuting someone who's confident. ''}\hspace{1em plus 1fill}---P10
    \end{quote}
    \begin{quote}
        \textit{``I was not sure if I should trust the Bard result\ldots but it sounds so smart, so it must be correct!''}\hspace{1em plus 1fill}---P63
    \end{quote}
\begin{enumerate}
    \item[(2)] \textbf{Be cautious about creating a conversational partner.} Participants attributed human characteristics to the AI (\nobreak \textit{Section \ref{discuss:3}}); anthropomorphizing can raise user \nobreak expectations \cite{Go2019} and cause overreliance \cite{Xiao2023}.
    \item[(3)] \textbf{Consider user customization.} Our findings suggest that access to the AI affect users of different expertise levels \nobreak differently. Through longitudinal exposure, the AI could build models of its human users and customize output by expertise and form a mutual theory of mind \cite{Ross2023, Wang2021}.
\end{enumerate}

\subsubsection{Cite sources.}\label{recs:2}
Participants preferred that Book resources were curated, deliberately written, peer-reviewed, and tried-and-true. When they navigated external search engines, they used websites that they were familiar with and looked for evidence of peer-review, such as highly ranked StackOverflow responses \cite{Abdalkareem2017}. In contrast, there was a sense of skepticism when using Bard; its generated output felt less intentional. Improving source attribution in LLMs and integrating appropriate citations into conversational output could bridge the perceived disconnect between generated and human-written output, which could lead to more appropriate trust \cite{Xiao2023}.
    \begin{quote}
        \textit{``There's a level of intentionality with the docs, like someone actually wrote this and put this together\ldots not knowing Bard, [its output] could be anything.''}\hspace{1em plus 1fill}---P75
    \end{quote}
    \begin{quote}
        \textit{``[I] tend to be skeptical about the answers people give without giving the source.'}\hspace{1em plus 1fill}---P78
    \end{quote}

\section{Limitations and future work}\label{limitations}
These findings may be limited to our particular context. Our study sample is limited to software engineers at a US-based technology company. They may have different attitudes towards AI and higher machine learning literacy as compared to laypeople, which may affect performance \cite{Chiang2022} and behaviors \cite{Gupta2021}. They've also been given direction to refrain from putting company-specific data in conversational AI systems, which may limit their familiarity.

Although empirical studies with other AI-based systems have produced similar findings on productivity and trust \cite{Gilson2023, Madi2023, Noy2023, Sun2022, Vai2022}, it's possible that our results are idiosyncratic to Bard. Reproduction of this task with other agents or a standardization of a behavioral task suite can help to generalize this work to other AI-based systems.

Finally, the task design may be scrutinized. Users interacted with the system for less than an hour. Perhaps trust formation takes longer time, more exposure, and more feedback; we may \nobreak benefit from a longitudinal study and extended follow-up \cite{Zhang2023}. This study was moderated, which may induce surveyor bias. Much of the recent, similar work on behavioral interactions with automated systems have been performed in unmoderated settings with online populations, who have been shown to behave differently \cite{Frechette2022}. Participants were given a flat thank-you gift regardless of performance on the exam; perhaps a higher-stakes setting or a piecewise incentive structure would induce more effort.

\section{Impact considerations}\label{Societal impact}
We demonstrate that users are willing to take up conversational AI to complete a workplace task, and that this assistance has the potential to improve user productivity. If users appropriately calibrate trust in these systems and use them in applicable settings, these systems can potentially increase productivity\cite{Hemmer2023}.

We also find that usage of these systems vary depending on user expertise. In this task, experts tended to distrust automated assistance. Novices, who were more likely to adopt and rely on these systems, were more susceptible to be influenced. This increased adoption of conversational AI by novices has the potential to equalize productivity across expertise. However, our findings suggest that adoption is not always beneficial: all participants exhibit automation complacency, and access to AI can be potentially detrimental in specific contexts. This could propagate inequitable outcomes \cite{Lund2023, Veinot2018}; learning differences may be exacerbated if conversational AI is applied in an academic setting, and malicious actors may employ these systems to disseminate disinformation to populations with lower literacy \cite{Wach2023}. 

Furthermore, the outputs of these systems hold weight. In our study, participants of all expertise levels could be convinced to change correct answers to incorrect answers following exposure to the AI. The impact of missing a few questions on a company-internal coding exam is fairly minimal. However, conversational AI has the capability to inform more consequential actions such as \nobreak obtaining a medical license \cite{Gilson2023}, consulting in clinical consultations \cite{Li2024}, or informing pandemic responses \cite{Xiao2023}. Furthermore, these LLM-based systems can mislead, hallucinate, and regurgitate incorrect information; for example, they can perpetuate unfair biases in the context of gender, race and religion \cite{Brown2020} and cite non-existent research \cite{Salvagno2023}. As we work to improve fairness and representation in these systems, we should concurrently improve model interpretability and user literacy to mitigate the potential of these systems to mislead.

\section{Conclusion}\label{conclusion}

\begin{table*}[ht]
\begin{tabular}{l}
\hline
\textbf{Results}                                                                                           \\
\begin{tabular}[c]{@{}l@{}}
1. \textit{Performance:} Access to the AI can improve performance on certain task types, and benefits novices more than experts.\end{tabular} \\
2. \textit{Efficiency:} Users may spend more time using the AI, yet perceive increased efficiency when using it. \\
3. \textit{Satisfaction: }Users may feel more productive using the AI, yet not more satisfied. \\
\begin{tabular}[c]{@{}l@{}}4.\textit{Trust:} Users may increase dependence on the AI over time. Users of all expertise levels are susceptible to mistrust.\end{tabular} \\
\begin{tabular}[c]{@{}l@{}}5. \textit{Distrust:} Relative to novices, experts are more likely to distrust the AI. This can punish their performance, \\ \hspace{.5cm}as experts are less likely to use the AI to recover from mistakes.\end{tabular}                 \\ \hline
\textbf{Behaviors}                                                                                         \\
1. Using the AI reduces search frictions, particularly for novices. \\
2. Users may reject the AI based on assumptions about its limitations. \\
3. Users do not perceive potential downsides of eliciting a second opinion. \\
4. Users exhibit confirmation bias and seek out agreement from the AI. \\
5. Users substitute effort to the AI, which reduces cognitive load.\\
6. Users exhibit automation complacency.\\
7. Users attribute blame asymmetrically.\\
8. Users perceive the AI as a collaborator.                                                           \\ \hline
\textbf{Recommendations}                                                                                   \\
1. Design for appropriate trust.                                                                      \\
2. Display the appropriate degree of confidence. \\
3. Be cautious about creating a conversational partner. \\
4. Consider user customization. \\
5. Cite sources.                             
\end{tabular}%
\caption{Summary of findings.}
\label{table:conclusions}
\end{table*}

In this study, we evaluated how access to conversational AI affects user productivity and trust formation through a user study of 76 software engineers as they completed an occupation-specific exam with and without access to a conversational AI agent. Broadly, we find that the effects on productivity and trust depend on the context and the user, and that having access to AI is not strictly better than not having access to AI. This evidence suggests that while these generative AI systems have the capability to affect and potentially augment worker productivity, they are not yet used in a way that can completely replace human effort or traditional resources.

We employed a mixed-methods approach of qualitative thematic analysis and quantitative statistical methods. This would at times yield seemingly inconsistent results; for example, participants perceived efficiency gains with AI assistance despite objectively taking more time on the task with AI assistance, and reported being less trustful of the AI despite increasingly depending on its outputs. We invite extensions of this work to continue exploring mixed-method approaches to capture a more holistic interpretation of behaviors.

As generative AI becomes more powerful and accessible, it becomes increasingly important for researchers and developers to understand the effect of these systems on users. We need to design these systems to account for human behaviors such as confirmation bias and automation complacency. System design should also prioritize minimizing the propagation of potentially misleading information, especially as our findings suggest that users, particularly novices, increasingly depend on these systems over time.

This paper contributes empirical evidence from a real-world scenario in the field of human-AI interaction \cite{Amershi2019}. We hope that this contribution will motivate more work in building more productive and trustworthy systems based on conversational AI.
\end{sloppypar}

\begin{acks}
We thank our pilot and study participants, Google's People and AI Research (PAIR) team, and the MIT Economics department for their valuable advice and support, notably Michael Terry, Lucas Dixon, Carrie Cai, Tobias Salz, Nikhil Agarwal, Ryan Mullins, Robert Upton, Martin Wattenberg, Fernanda Viégas, Hal Ableson, Mike Schaekermann, Key Lee, Frank Schilbach, Andrea Knight Dolan,  Mahima Pushkarna, and Hansa Srinivasan.
\end{acks}

\pagebreak
\bibliographystyle{ACM-Reference-Format}
\bibliography{References.bib}

\pagebreak
\appendix
\section{Survey Questions}\label{appendix:survey}
\subsection{Pre-Task Survey}\label{appendix:pre-survey}
In addition to the Trust in Automated Systems survey (\textit{Appendix \ref{appendix:jian}}), we asked the following to gauge background and self-described expertise. Many questions were adopted from a company-internal longitudinal survey on engineering productivity and satisfaction.

\textit{\textbf{Programming languages}}
\begin{enumerate}
    \item In the last three months, which languages have you used the most?
    \item Describe your level of familiarity in the following languages. (C++, Java, Python, Go, \ldots)
\end{enumerate}

\textit{\textbf{Knowledge resources}}
\begin{enumerate}
    \item In the past three months, how well have the following knowledge resources supported you in your development tasks? (Documentation, chat-based LLMs, forums, search tools, discussion spaces, \ldots)
\end{enumerate}

\textit{\textbf{Engineering satisfaction}}
\begin{enumerate}
    \item Overall, how satisfied are you with your experience as a developer at [company]?
    \item In the past three months, how productive have you felt at work at [company]?
    \item How often are you able to reach a high level of focus or achieve "flow" during development tasks?
    \item How satisfied are you with the quality of code that you produce?
    \item How satisfied are you with your engineering velocity?
\end{enumerate}

\textit{\textbf{LLM usage}}
\begin{enumerate}
    \item How often do you use LLM-assisted tools in your development tasks at work?
    \item If so, how well have LLM-assisted tools supported you in your development tasks in the past three months?
    \item Briefly describe any other ways LLM-assisted tools have supported you in your development tasks.
    \item Briefly describe any ways LLM-assisted tools have supported you in other tasks.
\end{enumerate}

\subsection{Post-Task Survey}\label{appendix:post-survey}
In addition to the Trust in Automated Systems survey (\textit{Appendix \ref{appendix:jian}}) and comparative satisfaction questions (\textit{Table \ref{table:sat}}), we asked the following to elicit free-form commentary.
\begin{enumerate}
    \item What was your approach for using non-LLM resources to answer questions?
    \item What was your approach for using Bard to answer questions?
    \item What was your approach for using non-LLM resources to verify your previous responses?
    \item What was your approach for using Bard to verify your previous responses?
    \item Which non-LLM resources did you use? How did they help?
    \item In the space below, please feel free to share any thoughts you have on the study.
\end{enumerate}


\section{Descriptive Statistics}\label{appendix:desc}
The following statistics about our sample are taken from both company-internal data and participant-reported pre-task survey responses.

\begin{figure*}
\begin{tabular}{cc}
  \includegraphics[width=70mm]{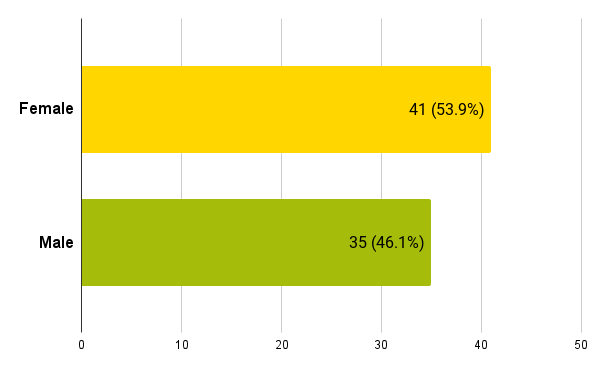} &   \includegraphics[width=70mm]{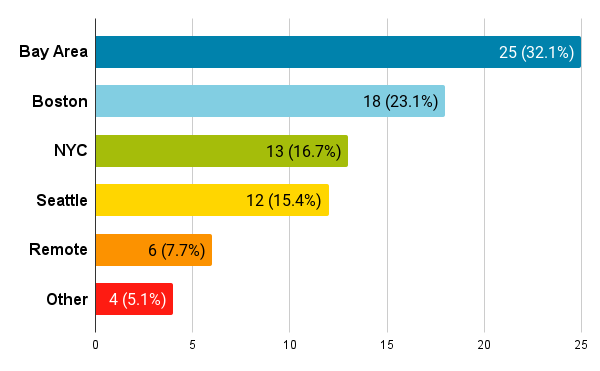} \\
a) Self-reported gender. & b) Job location. \\ \\
 \includegraphics[width=70mm]{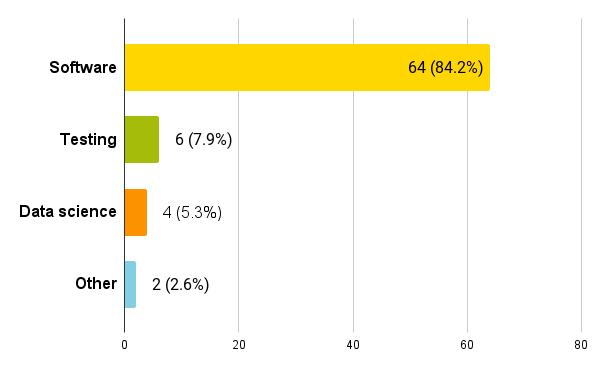} &   \includegraphics[width=70mm]{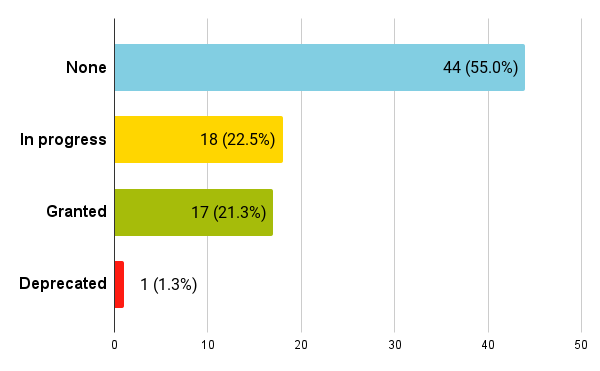} \\
c) Engineering ladder job role. & d) Java "readability" certification status.\footnotemark \\  \\
  \includegraphics[width=70mm]{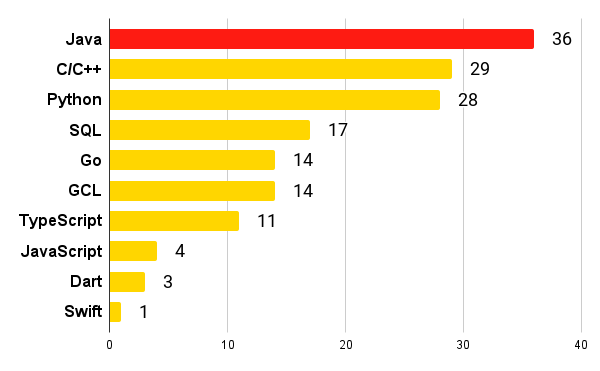} &   \includegraphics[width=70mm]{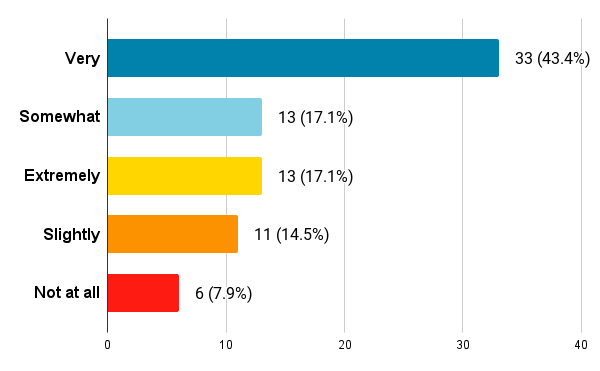} \\ 
e) Self-reported most common programming languages. & f) Self-reported familiarity with Java. \\
\end{tabular}
\label{fig:cat-stats}
\caption{Categorical statistics (n=76).}
\Description[Bar charts for descriptive statistics.]{53.9\% of the 76 participants identify as female. The majority of US participants are based in the Bay Area (32.1\%). 84.2\% of participants focus on software, but a few others focus on testing or data science. 55\% of our population have not obtained Java readability; 22.5\% are in the progress of obtaining this certification, and 21.3\% already have readability. Participants are most familiar with Java out of 10 programming languages. 43.4\% self-report that they are "very" familiar with Java.}
\end{figure*}

\setcounter{footnote}{15}\footnotetext{Those with \textit{In progress} or \textit{Granted} readability status have already taken this exam. Those with \textit{Deprecated} readability withdrew from the process of obtaining readability due to failure to take the exam.}

\section{Expertise Percentile Ranks}\label{appendix:reexp}

Expertise \cite{Krishna2001} is a multi-dimensional construct. \textit{Table \ref{table:exp-data}} shows summary statistics of expertise measures taken from company-internal data. To simplify the analysis, we constructed a joint objective measured expertise metric, weighting the following measures in decreasing order:
\begin{itemize}
    \item Java readability certification and status (\textit{Fig. 6d}), tiebreak by certifications in other languages
    \item Normalized number of Java changelists, tiebreak by changelists
 in other languages
    \item Most recent submitted Java, tiebreak by most recently submitted code in other languages
\end{itemize}

\begin{table*}\begin{tabular}{|lrrrr|}
\hline
\rowcolor[HTML]{C0C0C0} 
 &
  \multicolumn{1}{l}{\cellcolor[HTML]{C0C0C0}$\mu$} &
  \multicolumn{1}{l}{\cellcolor[HTML]{C0C0C0}$\sigma$} &
  \multicolumn{1}{l}{\cellcolor[HTML]{C0C0C0}Min} &
  \multicolumn{1}{l|}{\cellcolor[HTML]{C0C0C0}Max} \\ \hline
\multicolumn{5}{|l|}{\cellcolor[HTML]{EFEFEF}{
 \textit{Java expertise}}}                                                                \\ \hline
\multicolumn{1}{|l|}{\# lines of code (Java)}          & \multicolumn{1}{r|}{41,718}  & \multicolumn{1}{r|}{151,850} & \multicolumn{1}{r|}{0}   & 1,197,572 \\ \hline
\multicolumn{1}{|l|}{\# submitted changelists (Java)}  & \multicolumn{1}{r|}{157}     & \multicolumn{1}{r|}{}        & \multicolumn{1}{r|}{0}   & 2452      \\ \hline
\multicolumn{1}{|l|}{Most recent submitted Java} &
  \multicolumn{1}{r|}{\begin{tabular}[c]{@{}r@{}}October\\ 2022\end{tabular}} &
  \multicolumn{1}{r|}{-} &
  \multicolumn{1}{r|}{\begin{tabular}[c]{@{}r@{}}June\\ 2016\end{tabular}} &
  \begin{tabular}[c]{@{}r@{}}June\\ 2023\end{tabular} \\ \hline
\multicolumn{5}{|l|}{\cellcolor[HTML]{EFEFEF}\textit{Coding expertise}}                                                                                     \\ \hline
\multicolumn{1}{|l|}{\# lines of code (all languages)} & \multicolumn{1}{r|}{268,858} & \multicolumn{1}{r|}{680,043} & \multicolumn{1}{r|}{130} & 4,054,273 \\ \hline
\multicolumn{1}{|l|}{\# submitted changelists (all)}   & \multicolumn{1}{r|}{777}     & \multicolumn{1}{r|}{1306}    & \multicolumn{1}{r|}{9}   & 7132      \\ \hline
\multicolumn{1}{|l|}{Least recent submitted code} &
  \multicolumn{1}{r|}{\begin{tabular}[c]{@{}r@{}}October\\ 2019\end{tabular}} &
  \multicolumn{1}{r|}{-} &
  \multicolumn{1}{r|}{\begin{tabular}[c]{@{}r@{}}March\\ 2007\end{tabular}} &
  \begin{tabular}[c]{@{}r@{}}May\\ 2023\end{tabular} \\ \hline
\multicolumn{1}{|l|}{Most recent submitted code} &
  \multicolumn{1}{r|}{\begin{tabular}[c]{@{}r@{}}May\\ 2023\end{tabular}} &
  \multicolumn{1}{r|}{-} &
  \multicolumn{1}{r|}{\begin{tabular}[c]{@{}r@{}}September\\  2021\end{tabular}} &
  \begin{tabular}[c]{@{}r@{}}June\\ 2023\end{tabular} \\ \hline
\multicolumn{1}{|l|}{\# of readability certifications} & \multicolumn{1}{r|}{1}       & \multicolumn{1}{r|}{0.87}    & \multicolumn{1}{r|}{0}   & 4         \\ \hline
\end{tabular}%
\caption{Summary statistics on objective expertise scores.}
\label{table:exp-data}
\end{table*}

We similarly create a \textit{self-reported expertise} measure weighting self-reported Java experience (\textit{Fig. 9e, 9f}), engineering productivity (Appendix \ref{appendix:pre-survey}), and LLM expertise in decreasing order. These measures are then transformed into percentile ranks, which yields more robust estimates by reducing the influence of outliers \cite{Chetty2014}. The position of each individual’s expertise score in the distribution of scores is relative to all others in the primary analysis sample. Results are largely robust across most of the individual objective measures as well as the aggregated objective measure. Self-reported measures of expertise have low predictive power on the outcomes of interest. 


\section{Trust in Automated Systems Survey}\label{appendix:jian}
We administered Jian's Trust in Automated Systems survey before and after the task. Each statement is assessed on a 5-point, bipolar Likert scale with the possible options: \textit{Strongly disagree, Somewhat disagree, Neutral, Somewhat Agree, Strongly Agree.} The chart below shows frequencies of each option. To calculate the means and regression coefficients for the comparative satisfaction survey in \textit{Table \ref{table:sat}} with the same Likert responses, we map these options to numeric values [-2, -1, 0, 1, 2].

\begin{figure*}[ht]
    \includegraphics[width=10cm]{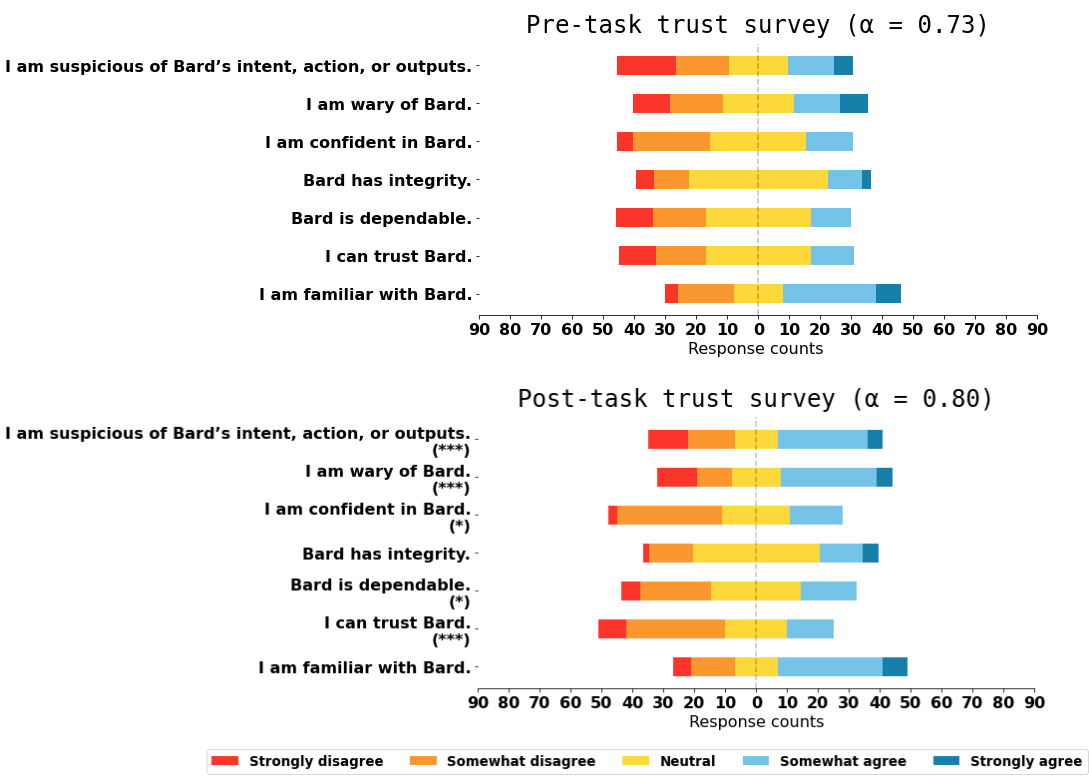}
    \caption{Pre- and post- task survey responses. $\alpha$ is Cronbach's alpha, a reliability measurement \cite{Tavakol2011}. Asterisks shows that the pre- and post- task distributions are significantly different, calculated using a $\chi^2$ independence test.}
    \label{fig:trust-change}
    \Description[A visualization of each statement in Jian's Trust in Automated Systems survey with the distribution of responses.]{A visualization of each statement in Jian's Trust in Automated Systems survey with the distribution of responses. The statements are "I am suspicious of Bard's intent, action or outputs," "I am wary of Bard," "I am confident in Bard," "Bard has integrity," "Bard is dependable," "I can trust Bard," and "I am familiar with Bard." Before the task, there are a lot of "Neutral" responses to these statements. Following the task, participants are more opinionated, stating more wary intent and less trust.}
\end{figure*}
\end{document}